\documentclass[twocolumn, a4paper]{revtex4}
\usepackage{graphicx}
\usepackage{dcolumn}
\usepackage{amsmath}
\usepackage{color}
\begin{document}
\hsize\textwidth\columnwidth\hsize\csname@twocolumnfalse\endcsname
\title{Observing sub-microsecond telegraph noise with the radio frequency single electron transistor}

\author{T. M. Buehler$^{1,2}$, D. J. Reilly$^{1,2}$, R. P. Starrett$^{1,2}$, V. C. Chan$^{1,3}$, A. R. Hamilton$^{1,2}$, A. S.
Dzurak$^{1,3}$ and R. G.Clark$^{1,2}$}

\affiliation{Centre for Quantum Computer Technology,
University of New South Wales, Sydney 2052, Australia}

\affiliation{Schools of Physics$^2$ and Electrical Engineering and Telecommunications$^3$, University of New South Wales,
Sydney 2052, Australia}

\begin{abstract}

\noindent 
Telegraph noise, which originates from the switching of charge between meta-stable trapping sites, becomes increasingly important as device sizes approach the nano-scale. 
For charge-based quantum computing, this noise may lead to decoherence and loss of read out fidelity. Here we use a radio frequency single electron transistor (rf-SET) 
to probe the telegraph noise present in a typical semiconductor-based quantum computer architecture. We frequently observe micro-second 
telegraph noise, which is a strong function of the local electrostatic potential defined by surface gate biases. We present a 
method for studying telegraph noise using the rf-SET and show results for a charge trap in which the capture and emission of a single
electron is controlled by the bias applied to a surface gate.  
\end{abstract}
\maketitle
The presence of $1/f$ noise, which is characterized by a spectral density inversely
proportional to frequency, has been observed in many diverse fields of physics \cite{dutta_horn1981}. In solid state 
devices the origin of this noise is generally believed to be the weighted sum of many independent noise
sources that physically correspond to fluctuating charge traps or defects. 
In the limit of strongly coupled charge traps, 
where the occupation of a {\it single} trap has a significant effect on 
the output signal, telegraph or switching noise is observed \cite{weisman}. In this instance the frequency 
dependence of the noise resembles a Lorentzian, deviating significantly from the usual $1/f$ form.  Owing to 
the importance of both $1/f$ and telegraph charge noise for the operation of nano-scale devices, there has been a recent 
resurgence of interest in this phenomena including its influence on decoherence rates in coherent quantum systems 
\cite{Itakura_PRB_03,Paladino_PRL_02, Martinis_PRB}. In particular, quantum computer (QC) architectures 
based on the charge degree of freedom \cite{Shnirman_PRL_97, Hayashi_PRL_03} are 
especially sensitive to charge noise, which has the potential to affect both phase coherence and qubit 
mixing times. 
\begin{figure}[t!]
\begin{center}
\includegraphics[width=8cm]{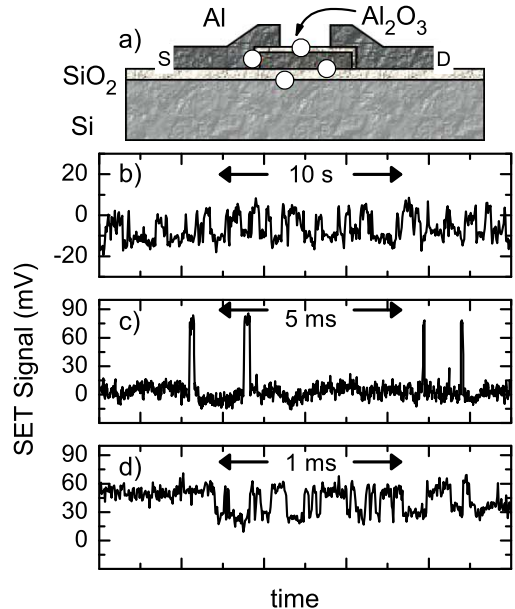}
\caption{\small{a) Device schematic showing the Al-SET (dark shading) and SiO$_2$/Si substrate. 
The white circles indicate regions likely to harbor charge traps. (b-d) rf-SET signal showing telegraph noise
as a function of time over several different time-scales.}}
\vspace{-0.5cm}
\end{center}
\end{figure}

Here we present charge noise measurements made on metal-oxide-semiconductor (MOS) devices cooled to mK
temperatures using a high bandwidth and sensitive electrometer: the radio frequency single electron transistor (rf-SET)
\cite{Schoelkopf_science_98}. The purpose of our work is to firstly study both the magnitude and frequency to which {\it telegraph} noise persists in
these structures and secondly, to investigate how it can be activated and controlled by the voltage biases applied to surface gates. Our method makes use of the rapid 
sample rate of the rf-SET detector to quickly locate regions of high telegraph switching in the parameter space defined by the gate voltage biases. Although the dependence 
of  telegraph noise on gate voltage has been extensively studied at lower frequencies (typically $<$1kHz) we note that the increased bandwidth of the 
rf-SET permits us to uncover large additional noise with switching times on the sub-microsecond scale \cite{Fujisawa_APL, Rice_Nature}.

Our interest is in devices that resemble architectures proposed for solid state QC, although these investigations also hold importance for classical nano-scale 
device operation. In particular our devices have a similar material structure to silicon-based QC proposals \cite{Kane_nature_98, Hollenberg_PRB_04} and consists of a high 
resistivity silicon substrate ( 5-7k$\Omega$cm) with a 5nm oxide, grown thermally at $T$ = $800\,^{\circ}\mathrm{C}$. The quality of the oxide and silicon oxide interface 
were
characterized using both capacitance profiling and by measuring the threshold voltage in a MOSFET configuration. 
Separate measurements indicate an interface trap density of $\sim 10^{12}/cm^2$ \cite{Mc_Camey}. Al-SETs and surface gates were fabricated on top of the oxide layer 
in a shadow mask process \cite{Fulton_PRL_87}. In contrast to conventional SETs, which are limited by the large shunting capacitance of the wiring, the rf-SET makes use of an 
impedance matching $LC$ network to transform the SET resistance towards the 50$\Omega$ characteristic impedance of a transmission line. In this mode rf-SETs have  
demonstrated bandwidths greater than 100MHz by mapping changes in resistance to the amount of reflected rf power from the $LCR$ network. The 
sensitivity of the rf-SET used in the experiments reported here (in the absence of telegraph noise) is $\delta q \sim 8 \mu e/\sqrt{Hz}$ at 1.1MHz (the details of 
our rf-setup can be found in 
\cite{Buehler_JAP}). 

As indicated by the white circles in Fig.1a, we identify several key locations  
within this device structure that are likely to harbor charge traps. With the exception of deep substrate traps, 
we believe that most trapping sites are located at the interface regions where the 
amorphous oxide forms bonds with either the silicon \cite{Uren_PRB_88} (below the SET) or aluminum (in the SET tunnel barriers) \cite{zorin}. 
\begin{figure}[t!]
\begin{center}
\includegraphics[width=8cm]{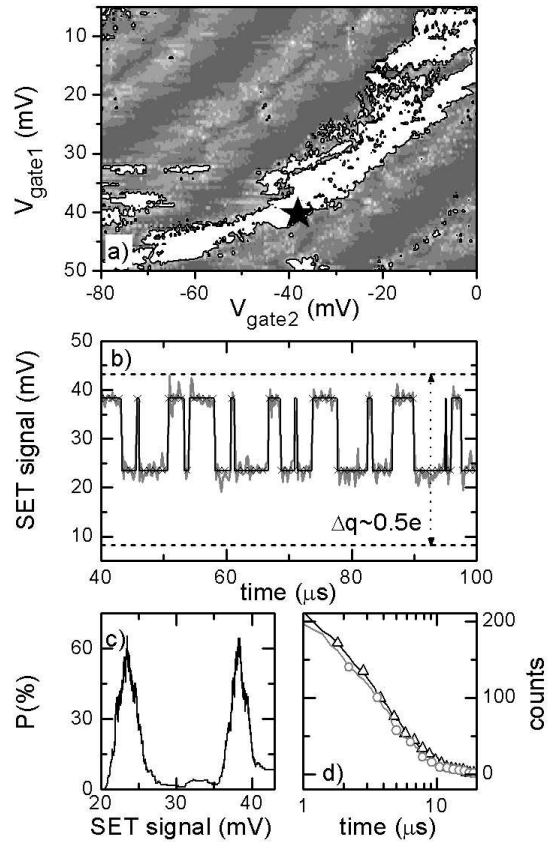}
\caption{\small{Mapping the dependence of telegraph noise on gate bias. At each gate bias configuration,  data is
taken for 2.0ms. a) Intensity plot showing the variance of each trace as a function of the gate bias.  Note the zone
moving diagonally across the plot that primarily corresponds to the switching of a two level fluctuator.
b) Data taken at the point indicated by the star in a). Dashed lines indicate the dynamic range of the SET. c) Shows signal histograms corresponding to the two states
of a two-level trap. d) Shows the trap capture and emission time histograms. For the case shown, equal time is spent in both the up (triangles) and down (circles)
states with a characteristic decay time of 3.4$\mu$s. }}\label{fig_intens} \vspace{-0.5cm}
\end{center}
\end{figure}

Turning now to our results, Figs.1b-d show the response of the rf-SET as a function of time on several different time-scales. Each trace is taken in succession, with some 
adjustment in offset charge between each trace. The data exhibits discontinuous switching of the rf-SET output signal corresponding to 
changes in induced charge on the SET island of order $\Delta q \sim $0.1$e$, with the exception of Fig.1c which is of order $\Delta q \sim $~0.2$e$. 
This telegraph noise is associated with charge capture and release events from traps in the surrounding
material that are strongly coupled to the SET. We see RTSs on all time-scales observable within the bandwidth of our detector ($\sim$10MHz in this case). As is 
evident 
by comparing data taken on different time-scales (e.g. Fig1.b \& c), the detector bandwidth is {\it crucial} in quantifying the amplitude of charge noise in these devices. 
In this way fast telegraph signals, such as those shown in Fig1.d would remain undetected by conventional dc SETs.

In order to study how the magnitude and switching time of the telegraph signals depends on the local electrostatic potential, we 
make use of the large bandwidth of the rf-SET to quickly vary the bias potential and
search for bias configurations that activate trap switching. Our characterization technique is outlined as follows. Firstly, 
biases applied to two surface gates
define the local potential. Secondly, at each gate bias the response of the rf-SET is monitored for 2.0 ms with a resolution of 200ns and the  average, variance, 
signal histogram and trap capture time are computed from the time-domain data. This procedure is repeated to cover a wide range of gate bias over which the degree of
charge noise may be mapped.  Fig.2a is an intensity plot of the variance of each  
2ms time-domain trace as a function of the two gate biases, taken on a different device to the data shown in Fig.1. The variance of the data (represented as an 
intensity) is a measure 
of the degree of telegraph noise present during the measurement time. 
The intensity plot reveals a distinct zone, mutually defined by each gate bias, where strong telegraph noise is evident (contoured region moving diagonally 
across the 
plot). In addition, the variation in sensitivity of the rf-SET can be seen as light regions also moving 
diagonally across the plot and corresponding to the edges of the Coulomb blockade oscillations. Although fast feedback techniques \cite{feedback}  can be used to eliminate 
such variations 
in sensitivity by 
maintaining a constant potential at the SET, 
we refrain from using such methods in order to study how telegraph noise is affected by changes in the potential near the SET.  
\begin{figure}[t!]
\begin{center}
\includegraphics[width=8cm]{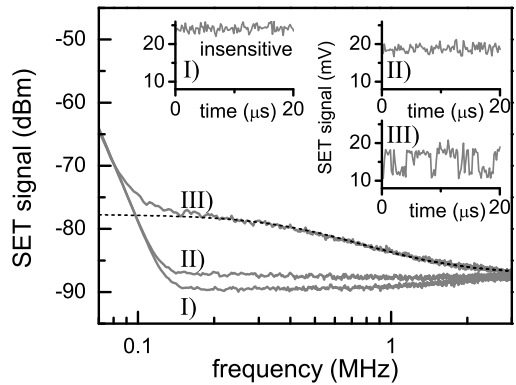}
\caption{\small{Frequency domain spectra and corresponding time-domain traces (insets) for different voltage biases applied to a nearby gate. For traces I) 
(and inset I) the rf-SET is in blockade and insensitive to charge noise (only intrinsic noise shown). 
For trace II) (and inset II) the data was taken with the gate bias set to condition 1 (see main text). For trace III) (and inset III) the gate bias set to condition 
2. For III) the dashed line is a Lorentzian fitted to the spectra. }}
\label{fig_rts_spec} \vspace{-0.5cm}
\end{center}
\end{figure}

Figs.2b-d are examples of the information that can be extracted at each point in the intensity plot. Fig.3b is a 100$\mu$s long segment of the 2ms time-domain 
data taken at the bias point indicated by the star in Fig.2a. The data shows telegraph switching between two trapping sites on microsecond time-scales, with the dynamic 
range 
of the SET ($\sim$0.5e) indicated by the dashed lines. Fig.2c and 2d show  signal probability histograms of the two distinct states  and trap capture- 
and emission-time histograms respectively, computed from a fit to the  2ms of time-domain data (the fit is shown in 
Fig.2b). For the fluctuator examined here an electron spends equal time (decay time of 3.4$\mu$s) in both the up and down trapped states. Of further interest the diagonal 
zone in Fig.2a where a high degree of switching is evident follows a Coloumb blockade peak edge, where the potential at the SET is kept constant. This suggests that the 
trap we observe here is located in or very close to the SET.   

\begin{figure}[t!]
\begin{center}
\includegraphics[width=8cm]{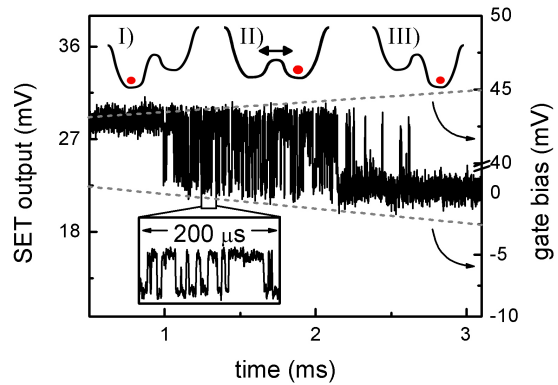}
\caption{\small{rf-SET signal (left y-axis) as a function of time as a differential gate bias (right y-axis) is simultaneously applied.  The data shows
that the occupancy of a charge trap can be controlled by a gate bias and detected on fast time-scales. I-III) are schematics
corresponding to the trap configuration as a function of time. }}\label{fig_rts_switch} \vspace{-0.5cm}
\end{center}
\end{figure}

Having presented our technique for mapping telegraph noise we now turn to discuss the presence of sub-microsecond switching.
Fig.3 compares the SET output spectra (obtained with a spectrum analyzer over several minutes) at
different gate biases (data taken from the same device as data in Fig.2). For the case where the
SET is in blockade (spectra and time trace I)) the data represents the intrinsic system
noise and {\it not} the input charge noise detected directly by the SET electrometer.
For frequencies $<$ 100kHz, there is little difference between the traces as the output is dominated by
the intrinsic $1/f$ noise of the measurement system. Above 100kHz however, we find a significant increase in charge noise
between spectra II) and III), which were obtained at two different (but near equally sensitive) gate bias settings.
The black spectra is clearly associated with rapid telegraph noise of order $\sim 0.1e$ (see inset III)) and fits
the high frequency `tail' of the Lorentzian distribution (dashed line is a Lorentzian fit) expected for a two level fluctuator strongly coupled to the SET 
\cite{Fujisawa_APL,weisman}.
Under this condition the spectral density of the noise exceeds the intrinsic shot and thermal noise of the detector for
frequencies up to $\sim$2MHz, well above the typical corner frequency of 10-100kHz generally associated with 1/f charge noise. Although we present data here for a 
fluctuator  that exhibits {\it microsecond} switching we have also observed switching on faster time-scales (up to 70ns - the bandwidth of our rf-SET), consistent 
with counting errors seen in experiments on single electron pumps \cite{Kautz_PRB_00}.

Finally we present data in which the state of a two-level charge trap can be reversibly {\it switched} by the bias applied to surface gates \cite{Cobden_PRL_93}. Here we 
study the same device from which data was taken for  Figs.2 \& 3. 
Fig.4 shows the rf-SET signal (left y-axis) while a differential 
voltage ramp is applied between two gates (right y-axis). When the differential bias is small the electron is predominately in the up (left) state, 
as shown schematically in inset I). As the bias increases, occupation of the trap 
fluctuates between the two equilibrium levels until the electron is captured and maintained in the down (right) state for increased differential
gate bias as shown in inset III). The differential bias changes by only $\Delta V$=3.8mV across the plot. In contrast to the measurements presented in Fig.2, the results 
shown here were taken using a compensation  technique 
in which the potential at the SET is held constant while the potential away from the SET is varied. As a result we believe that this two-level 
trap is located in either the substrate or at the silicon-oxide interface away from the SET island, where the trap can be switched as the potential changes 
with the differential gate bias.  Of further note, no qualitative change in behavior of the trap was observed with increasing temperature to $T$=500mK (or 
$B$-field to 
2T), consistent with the notion that the fluctuations are likely associated with {\it tunneling} between two trapping sites (as indicated in the insets) and not 
over-barrier thermal activation \cite{Scofield_APL_2000}.

Taken together with the characterization technique shown in Fig.2, the ability to 
find and control charge traps on fast time-scales, opens up the possibility of studying the noise properties of {\it single} two-level fluctuators, including their 
quantum dynamics \cite{Xiao_PRL_03}. Although
here we have only shown preliminary data, these results demonstrate the capability of the rf-SET to investigate telegraph noise in nano-scale devices on time-scales 
previously inaccessible. Future efforts will focus on combining the charge mapping technique presented here with a source of microwave photons to provide insight into 
the mechanisms driving fluctuations on these time-scales. In particular the interplay of microwave photon assisted tunneling and thermal activation of charge 
switching will be crucial to the study and control of decoherence in quantum systems.     

In conclusion, we have shown that telegraph noise in Silicon MOS devices, occurs on all time-scales observable within the 
bandwidth of our rf-SET. We have presented a technique for the characterization of this noise as a function of the local potential and demonstrated gate-bias controlled switching of 
a single charge trap on fast time-scales, an ideal platform in which to explore the feasibility of charge based quantum computation.

We thank D. Barber for technical support. This work was supported
by the  Australian Research Council, the Australian Government and
by the US National Security Agency (NSA), Advanced Research and
Development Activity (ARDA) and the Army Research Office (ARO)
under contract number DAAD19-01-1-0653. DJR acknowledges a  
Hewlett-Packard Fellowship.


\begin{thebibliography}{22}
\small
\expandafter\ifx\csname natexlab\endcsname\relax\def\natexlab#1{#1}\fi
\expandafter\ifx\csname bibnamefont\endcsname\relax
  \def\bibnamefont#1{#1}\fi
\expandafter\ifx\csname bibfnamefont\endcsname\relax
  \def\bibfnamefont#1{#1}\fi
\expandafter\ifx\csname citenamefont\endcsname\relax
  \def\citenamefont#1{#1}\fi
\expandafter\ifx\csname url\endcsname\relax
  \def\url#1{\texttt{#1}}\fi
\expandafter\ifx\csname urlprefix\endcsname\relax\def\urlprefix{URL }\fi
\providecommand{\bibinfo}[2]{#2}
\providecommand{\eprint}[2][]{\url{#2}}

\bibitem[{\citenamefont{Dutta and Horn}(1981)}]{dutta_horn1981}
\bibinfo{author}{\bibfnamefont{P.}~\bibnamefont{Dutta}} \bibnamefont{and}
  \bibinfo{author}{\bibfnamefont{P.~M.} \bibnamefont{Horn}},
  \bibinfo{journal}{Rev. Mod. Phys.} \textbf{\bibinfo{volume}{53}},
  \bibinfo{pages}{497} (\bibinfo{year}{1981}).

\bibitem[{\citenamefont{Weissman}(1988)}]{weisman}
\bibinfo{author}{\bibfnamefont{M.~B.} \bibnamefont{Weissman}},
  \bibinfo{journal}{Rev. Mod. Phys.} \textbf{\bibinfo{volume}{60}},
  \bibinfo{pages}{537} (\bibinfo{year}{1988}).

\bibitem[{\citenamefont{Itakura and Tokura}(2003)}]{Itakura_PRB_03}
\bibinfo{author}{\bibfnamefont{T.}~\bibnamefont{Itakura}} \bibnamefont{and}
  \bibinfo{author}{\bibfnamefont{Y.}~\bibnamefont{Tokura}},
  \bibinfo{journal}{Phys. Rev. B.} \textbf{\bibinfo{volume}{67}},
  \bibinfo{pages}{195320} (\bibinfo{year}{2003}).

\bibitem[{\citenamefont{Paladino et~al.}(2002)\citenamefont{Paladino, Faoro,
  Falci, and Fazio}}]{Paladino_PRL_02}
\bibinfo{author}{\bibfnamefont{E.}~\bibnamefont{Paladino {\it et al.}}},
  \bibinfo{journal}{Phys. Rev. Lett.} \textbf{\bibinfo{volume}{88}},
  \bibinfo{pages}{228304} (\bibinfo{year}{2002}).

\bibitem[{\citenamefont{Martinis et~al.}(2003)\citenamefont{Martinis, Nam,
  Aumentado, Lang, and Urbina}}]{Martinis_PRB}
\bibinfo{author}{\bibfnamefont{J.~M.} \bibnamefont{Martinis {\it et al.}}},
  \bibinfo{journal}{Phys. Rev. B.} \textbf{\bibinfo{volume}{67}},
  \bibinfo{pages}{094510} (\bibinfo{year}{2003}).

\bibitem[{\citenamefont{Shnirman et~al.}(1997)\citenamefont{Shnirman, Schoen,
  and Hermon}}]{Shnirman_PRL_97}
\bibinfo{author}{\bibfnamefont{A.}~\bibnamefont{Shnirman}},
  \bibinfo{author}{\bibfnamefont{G.}~\bibnamefont{Schoen}}, \bibnamefont{and}
  \bibinfo{author}{\bibfnamefont{Z.}~\bibnamefont{Hermon}},
  \bibinfo{journal}{Phys. Rev. Lett.} \textbf{\bibinfo{volume}{79}},
  \bibinfo{pages}{2371} (\bibinfo{year}{1997}).

\bibitem[{\citenamefont{Hayashi et~al.}(2003)\citenamefont{Hayashi, Fujisawa,
  Cheong, Jeong, and Hirayama}}]{Hayashi_PRL_03}
\bibinfo{author}{\bibfnamefont{T.}~\bibnamefont{Hayashi {\it et al.}}},
  \bibinfo{journal}{Phys. Rev. Lett.} \textbf{\bibinfo{volume}{91}},
  \bibinfo{pages}{226804} (\bibinfo{year}{2003}).

\bibitem[{\citenamefont{Schoelkopf et~al.}(1998)\citenamefont{Schoelkopf,
  Wahlgren, Kozhevnikov, Delsing, and Prober}}]{Schoelkopf_science_98}
\bibinfo{author}{\bibfnamefont{R.~J.} \bibnamefont{Schoelkopf {\it et al.}}},
  \bibinfo{journal}{Science} \textbf{\bibinfo{volume}{280}},
  \bibinfo{pages}{1238} (\bibinfo{year}{1998}).

\bibitem[{\citenamefont{Fujisawa and Hirayama}(2000)}]{Fujisawa_APL}
\bibinfo{author}{\bibfnamefont{T.}~\bibnamefont{Fujisawa}} \bibnamefont{and} 
  \bibinfo{author}{\bibfnamefont{Y.}~\bibnamefont{Hirayama}},
  \bibinfo{journal}{App. Phys. Lett.} \textbf{\bibinfo{volume}{77}},
  \bibinfo{pages}{543} (\bibinfo{year}{2000}).

\bibitem[{\citenamefont{Lu et~al.}(2003)\citenamefont{Lu, Ji, Pfeiffer, West,
  and Rimberg}}]{Rice_Nature}
\bibinfo{author}{\bibfnamefont{W.}~\bibnamefont{Lu {\it et al.}}},
  \bibinfo{journal}{Nature} \textbf{\bibinfo{volume}{423}},   
  \bibinfo{pages}{422} (\bibinfo{year}{2003}).

\bibitem[{\citenamefont{Kane}(1998)}]{Kane_nature_98}
\bibinfo{author}{\bibfnamefont{B.~E.} \bibnamefont{Kane}},
  \bibinfo{journal}{Nature} \textbf{\bibinfo{volume}{393}},
  \bibinfo{pages}{133} (\bibinfo{year}{1998}).
  
\bibitem[{\citenamefont{Hollenberg et~al.}(2004)\citenamefont{Hollenberg,
  Dzurak, Wellard, Hamilton, Reilly, Milburn, and Clark}}]{Hollenberg_PRB_04}
\bibinfo{author}{\bibfnamefont{L.~C.~L.} \bibnamefont{Hollenberg {\it et al.}}},
  \bibinfo{journal}{Phys. Rev. B.} \textbf{\bibinfo{volume}{69}},
  \bibinfo{pages}{113301} (\bibinfo{year}{2004}).

\bibitem[{\citenamefont{Camey}(2004)}]{Mc_Camey}
\bibinfo{author}{\bibfnamefont{D.~M.} \bibnamefont{Camey {\it et al.}}},
  \bibinfo{journal}{Unpublished}  (\bibinfo{year}{2004}).

\bibitem[{\citenamefont{Fulton and Dolan}(1987)}]{Fulton_PRL_87}
\bibinfo{author}{\bibfnamefont{T.~A.} \bibnamefont{Fulton}} \bibnamefont{and}
  \bibinfo{author}{\bibfnamefont{G.~J.} \bibnamefont{Dolan}},
  \bibinfo{journal}{Phys. Rev. Lett.} \textbf{\bibinfo{volume}{59}},
  \bibinfo{pages}{109} (\bibinfo{year}{1987}).
  
\bibitem[{\citenamefont{Buehler and et~al}(2003)}]{Buehler_JAP}
\bibinfo{author}{\bibfnamefont{T.~M.} \bibnamefont{Buehler {\it et al.}}} 
\bibinfo{journal}{To appear in J. App. Phys. (arXiv:cond-mat/0302085.)}  
(\bibinfo{year}{2003}).

\bibitem[{\citenamefont{Uren et~al.}(1988)\citenamefont{Uren, Kirton, and
  Collins}}]{Uren_PRB_88}
\bibinfo{author}{\bibfnamefont{M.~J.} \bibnamefont{Uren {\it et al.}}},
  \bibinfo{journal}{Phys. Rev. B.} \textbf{\bibinfo{volume}{37}},
  \bibinfo{pages}{8346} (\bibinfo{year}{1988}).

\bibitem[{\citenamefont{Zorin et~al.}(1996)\citenamefont{Zorin, Ahlers,
  Niemeyer, Weimann, , Wolf, Krupenin, and Lotkhov}}]{zorin}   
\bibinfo{author}{\bibfnamefont{A.~B.} \bibnamefont{Zorin {\it et al.}}},
 \bibinfo{journal}{Phys. Rev. B}
  \textbf{\bibinfo{volume}{53}}, \bibinfo{pages}{13682} (\bibinfo{year}{1996}).

\bibitem[{\citenamefont{Segall and et~al.}(2002)}]{feedback} 
\bibinfo{author}{\bibfnamefont{K.}~\bibnamefont{Segall {\it et al.}}} 
\bibinfo{journal}{App. Phys. Lett.}
  \textbf{\bibinfo{volume}{81}}, \bibinfo{pages}{4859} (\bibinfo{year}{2002}).

\bibitem[{\citenamefont{Kautz et~al.}(2000)\citenamefont{Kautz, Keller, and   
  Martinis}}]{Kautz_PRB_00}
\bibinfo{author}{\bibfnamefont{R.~L.} \bibnamefont{Kautz {\it et al.}}},
\bibinfo{journal}{Phys. Rev. B}
  \textbf{\bibinfo{volume}{62}}, \bibinfo{pages}{15888} (\bibinfo{year}{2000}).
  
\bibitem[{\citenamefont{Cobden et~al.}(1993)\citenamefont{Cobden, Uren, and   
  Pepper}}]{Cobden_PRL_93}
\bibinfo{author}{\bibfnamefont{D.~H.} \bibnamefont{Cobden {\it et al.}}},
  \bibinfo{journal}{Phys. Rev. Lett.} \textbf{\bibinfo{volume}{71}},
  \bibinfo{pages}{4230} (\bibinfo{year}{1993}).

\bibitem[{\citenamefont{Scofield et~al.}(2000)\citenamefont{Scofield, Borland,
  and Fleetwood}}]{Scofield_APL_2000}
\bibinfo{author}{\bibfnamefont{J.~H.} \bibnamefont{Scofield {\it et al.}}},
  \bibinfo{journal}{Appl. Phys. Lett.} \textbf{\bibinfo{volume}{76}},
  \bibinfo{pages}{3248} (\bibinfo{year}{2000}).

\bibitem[{\citenamefont{Xiao et~al.}(2003)\citenamefont{Xiao, Martin, and
  Jiang}}]{Xiao_PRL_03}
\bibinfo{author}{\bibfnamefont{M.}~\bibnamefont{Xiao {\it et al.}}},
  \bibinfo{journal}{Phys. Rev. Lett.} \textbf{\bibinfo{volume}{91}},
  \bibinfo{pages}{078301} (\bibinfo{year}{2003}).

\end{thebibliography}
\end{document}